\documentclass[10pt,letterpaper]{article} 

\usepackage{olArxiv}
\usepackage[draft]{hyperref}
\usepackage{amsmath}

\begin{document}


\title{A 100~mW monolithic Yb waveguide laser fabricated using the femtosecond laser direct-write technique}


\author{Martin~Ams,$^{*}$ Peter~Dekker, Graham~D.~Marshall, and Michael~J.~Withford}

\address{
MQ Photonics Research Centre\\
Centre for Ultrahigh bandwidth Devices for Optical Systems (CUDOS)\\
Department of Physics, Macquarie University, NSW 2109, Australia\\
$^*$Corresponding author: mams@physics.mq.edu.au
}

\begin{abstract}A femtosecond laser-written monolithic waveguide laser (WGL) oscillator based on a distributed feedback (DFB) architecture and fabricated in ytterbium doped phosphate glass is reported. The device lased at 1033~nm with an output power of 102~mW and a bandwidth less than 2~pm when bidirectionally pumped at 976~nm. The WGL device was stable and operated for 50 hours without degradation. This demonstration of a high performance WGL opens the possibility for creating a variety of narrow-linewidth laser designs in bulk glasses.\\\end{abstract}


\newcommand{\intensity}[2]{#1$\times$10$^{#2}$~\textrm{W/cm$^2$}}

\newcommand{\dist}[1]{#1~$\mu$m}

\newcommand{\ri}[2]{#1$\times$10$^{#2}$}

\newcommand{\loss}[1]{#1~dB/cm}

\newcommand{\speed}[1]{#1~$\mu$m/s}

\newcommand{\energy}[1]{#1~$\mu$J}

\newcommand{\obj}[2]{#1$\times$ (#2 NA)}

\noindent There is an ever-growing need for integrated optical devices for use in guided wave applications such as communications and sensing. International activity over the past 40 years has led to various technologies being used for the fabrication of such devices: ion exchange~\cite{nikonorov1999}, silicon-on-insulator~\cite{eldada2004} and silica-on-silicon\cite{eldada2001}. A relatively new fabrication method that also shows good promise in this field is that of ultrafast laser direct-writing. It was shown in 1996 that focussed femtosecond laser pulses can induce a permanent refractive index change in dielectric media~\cite{davis1996}. The index change is initiated by the electrons in the material absorbing energy from the radiation field via various nonlinear mechanisms. Because the absorption is nonlinear, material modification remains localised at the focal point where the laser intensity is highest. Hence, by translating the material through the focus of a femtosecond laser beam, a pathway of refractive index change can be produced using this technique. In many glasses femtosecond laser irradiation results in an increase in refractive index and hence waveguiding, while in crystalline hosts such as YAG and LiNbO$_3$, waveguiding can only be achieved through suppressed cladding arrangements or induced stress fields as the index change is typically negative~\cite{okhrimchuk2005}. By varying the writing geometry, the material and femtosecond laser properties, 2D and 3D optical waveguide devices with different characteristics can be fabricated in the bulk of many transparent materials~\cite{gattass2008}.

Ultrafast laser writing has been successfully applied to active materials resulting in the creation of waveguide amplifiers and waveguide lasers (WGLs)~\cite{kawamura2004, okhrimchuk2005, dellavalle2007, psaila2008, marshall2008}. For example, a femtosecond laser encoded distributed feedback (DFB) colour center laser was reported in a LiF crystal operating in a pulsed mode at 704~nm using an optical parametric oscillator (OPO) as a pump~\cite{kawamura2004}. However, the grating structure of this device was incoherent, the linewidth was significantly broader than that normally expected of DFB lasers and no output powers were reported. More recently, bulk glass WGL devices have been demonstrated using external fiber Bragg gratings (FBGs) to complete a cavity around a rare-earth doped waveguide amplifier~\cite{dellavalle2007, psaila2008}. This configuration is a bulk glass waveguide analogue to a fiber laser and allows for efficient pumping with standard fiber pigtailed diode lasers. Output powers up to 80~mW with slope efficiencies of 21\%, at 1.5~$\mu$m using Er/Yb co-doped glass, have been reported~\cite{dellavalle2007}.

To make a truly monolithic device, we previously reported on the fabrication of a C-band WGL constructed from a waveguide-Bragg grating (WBG) acting as a DFB resonator~\cite{marshall2008}. While being the first demonstration of a new class of laser, this device had a modest output power of the order of 1~mW, a low slope efficiency and only operated slightly above threshold which made device characterisation difficult. An alternative to the 1.5~$\mu$m Er/Yb system~\cite{marshall2008} is the high gain Yb only system operating around 1~$\mu$m. Trivalent Yb offers a lower quantum defect than Er which reduces phonon induced heating in the grating structure thus lessening pump induced thermal chirp of the WBG. In this Letter, we report a Yb WGL with a maximum single-end output power of 102~mW at 1032.59~nm. The pump power threshold was approximately 115~mW and the optical efficiency was over 17\%. This device represents what we believe is the first demonstration of an efficient high power monolithic WGL created entirely using the femtosecond laser direct-write technique.


The femtosecond laser direct-write technique was used to create both the waveguide and the WBG simultaneously and in a single processing step. The laser used to fabricate this device was a 1~kHz repetition rate, 120~fs pulse length, 800~nm regeneratively amplified Ti:sapphire laser that was focused into the glass sample using a 20$\times$ microscope objective. The writing-laser beam was circularly polarised to minimize waveguide propagation losses~\cite{ams2006} and focussed at a depth of 170~$\mu$m below the surface of the substrate. A \dist{520} width slit was placed in front of the focussing objective to create waveguides with a circularly symmetric cross-section~\cite{ams2005}. The glass sample was translated at 25~$\mu$m/s through the focussed writing beam that was 100\% intensity modulated at approximately 75~Hz and with a 50:50 mark–space ratio to create the required 335~nm period refractive index perturbation that corresponded to a first order WBG. This method of intensity modulation to create a WBG is similar to that reported in~\cite{zhang2007a} and further details pertaining to device fabrication can be found in~\cite{marshall2008}.

The substrate used was a ``QX'' phosphate glass host (Kigre Inc., USA) doped with 9\% (by weight) ytterbium. The peak in absorption was at 974.5~nm (absorption coefficient 10~cm$^{-1}$) and the material exhibited a gain bandwidth of approximately 80~nm centered around 1030~nm. The waveguides had a typical physical diameter of \dist{8} and guided a single transverse mode with a 1/e$^2$ diameter of \dist{13}. Refractive index profilometry (Rinck Electronik, measured at 633~nm) indicated the peak refractive index contrast of the waveguide was \ri{1.4}{-3}. Test WBGs written at 1535~nm (outside the absorption band of Yb) were used to provide an upper bound of the grating strength resulting in a $\kappa L$ of 2.1 for a WBG of length 9.5~mm. From this value, the refractive index contrast between the grating periods was calculated to be \ri{1.1}{-4} indicating that the induced WBG contrast was approximately 8\% of the total refractive index change between the waveguide and the bulk material.

\begin{figure}[htb]
\centerline{\includegraphics[width=8.3cm]{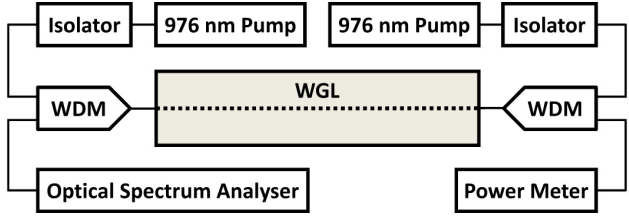}}
\caption{Schematic of the bidirectional WGL pumping and diagnostic setup.}
\label{setup}
\end{figure}

Pumping of the waveguide was achieved using single mode optical fibers and wavelength division multiplexors (WDMs) to deliver a combined total pump power of up to 726~mW at 976~nm. Laser output was collected from both ends of the WGL as shown in Fig.~\ref{setup}. The position of the OSA and power meter could be interchanged to compare the laser output from each end. To better match the \dist{13} mode field diameter (MFD) of the WGL to the MFD of the pump/collection fibers, short sections of graded index optical fiber (GIF625) were fusion spliced to the fiber tips to act as mode field convertors~\cite{yablon2005}.

Initially, the prepared WBG length was 23.2~mm and, when bidirectionally pumped, each end was found to lase independently of the other indicating that the sample length was longer than that suitable for the available double-end pump power. This was further reinforced by the fact that there was little or no cooperative luminescence (characteristic of a well pumped material) emanating from the center of the WGL. Even so, pumping the WGL and collecting the laser light from a single-end resulted in a pump threshold of 116~mW and a maximum output power of 38~mW (using a single-end pump power of 400~mW).

\begin{figure}[htb]
\centerline{\includegraphics[width=8.3cm]{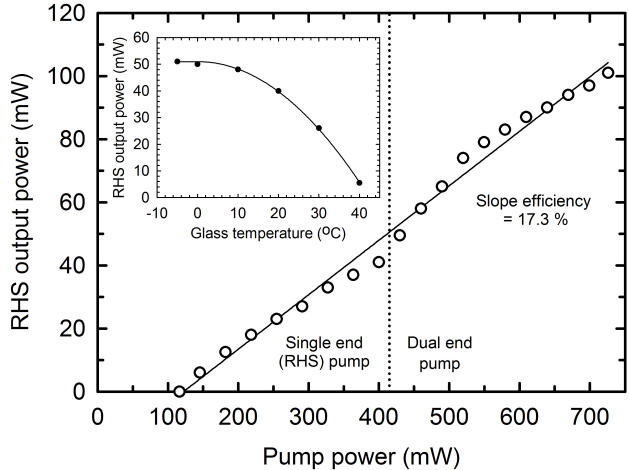}}
\caption{Laser output power, at room temperature, as a function of pump power for single and dual end pump geometries using a 9.5~mm long WGL. Inset shows the effect of bulk glass temperature on laser output power.}
\label{slope}
\end{figure}

To enable effective double-end pumping the WBG was cut back to a length of 9.5~mm thereby creating a device with an original facet on the left hand side (LHS) and a new facet on the right hand side (RHS).
The single-end pump power thresholds and output powers of the original uncut sample and of the 9.5~mm WGL were nearly identical. This indicated that the \textit{effective} single-end pumped laser length had not changed. For the short 9.5~mm device the forward propagating WGL power was 17~mW at the maximum single-end pump power. The reverse propagating output power as a function of incident pump power for the 9.5~mm long WGL is shown in Fig.~\ref{slope}.
The output power increased by up to 25\% by reducing the bulk glass temperature to approximately 20$^\circ$C below ambient as shown in the inset of Fig.~\ref{slope}.
The increased output power at this wavelength is characteristic of Yb in glass which operates in the 3-level regime with moderate ground state thermal depopulation.

Using the maximum amount of available bidirectional pump power we obtained 102~mW of output at 1032.59~nm with a slope efficiency of 17.3\% from the RHS laser facet. The output from the LHS of the WGL decreased from the single-end pumped value of 17~mW to 2~mW when dual end pumped. Given that the WGL was designed to be symmetric, a simplistic interpretation of the device architecture would have equal powers emanating from each facet. However, small perturbations to the period and phase of the DFB structure are capable of producing near unidirectional DFB laser operation~\cite{yelen2004}. The asymmetric output behaviour of the single and dual end pumped WGLs is characteristic of the refractive index chirp known to exist in the laser host material and the pump induced thermal gradients that also affect the refractive index in the WBG. These two parameters combine to produce an effective phase shift in the WBG that is dependent on the pump power intensity and distribution.

\begin{figure}[htb]
\centerline{\includegraphics[width=8.3cm]{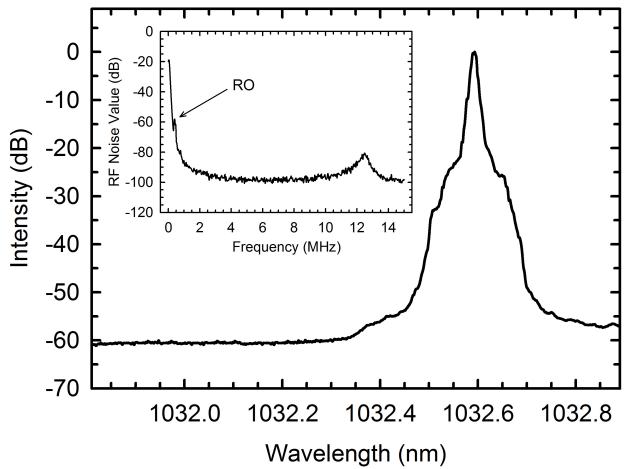}}
\caption{Output spectrum of the WGL at 102~mW output power. The ordinate axis scale is referenced to the peak. Inset shows the RF noise spectrum of the WGL when pumped with linearly polarized light.}
\label{osa}
\end{figure}

The output spectrum of the WGL operating at maximum input power is shown in Fig.~\ref{osa}. The shape of the laser spectrum was the same when viewed from either end. The WGL output spectrum is dominated by a single peak at 1032.59~nm and has a 3~dB full width that was limited by the 10~pm instrument resolution. Measurements of the WGL output on a scanning Michelson interferometer wavemeter indicated an upper bound of the linewidth of 2~pm. The side mode suppression ratio of the WGL is at least 20~dB and the laser output is accompanied by weak ASE that falls within the estimated grating reflection bandwidth of 300~pm. The small perturbations to the otherwise smooth ASE profile are thought to be caused by the internal Fabry Perot resonances between the uncoated WGL facets.

The WGL device characteristics were stable over the 50 hour period of characterisation demonstrating that the WBG is permanent. Measurements of the laser noise were made using an oscilloscope and radio frequency (RF) analyser coupled to a 1~GHz bandwidth photodiode. Pumping with unpolarised light we typically obtained lasing on two frequency degenerate polarization modes (within the 2~pm linewidth) with polarizations parallel and orthogonal to the direction of the WBG writing beam. The degeneracy of these laser polarization modes indicates that the birefringence of the WBG is less than \ri{2}{-6}. Strong gain competition between the polarization modes was observed and often resulted in fluctuations in the laser output intensity. Removal of the non-polarization maintaining (non-PM) WDMs and using PM fiber enabled us to pump with polarizations aligned to the birefringent axes of the WBG resulting in a single linearly polarized output mode aligned in the same direction as the pump polarization. In this case the fluctuations in laser output were suppressed and measurements of the laser noise spectrum indicated a clear relaxation oscillation (RO) peak at 480~kHz and a single unexplained noise peak at 12.5~MHz as shown in the inset of Fig.~\ref{osa}. We were unable to take measurements of the output power in this condition.

The femtosecond laser direct-write technique provides an approach to fabricating WBGs with arbitrary Bragg wavelengths. As an example of this we fabricated a WGL at 1044~nm in the same block of glass using identical writing parameters to those for the WGL outlined above. In this case we obtained, from single-end pumping, a maximum counter-propagating laser output power of 20~mW. This result was not further optimized for device length or pumping conditions.

In this Letter we have demonstrated an efficient, high power WGL operating at 1032.59~nm in Yb-doped phosphate glass with single ended output powers over 100~mW. The use of the femtosecond laser direct-write technique enabled us to fabricate narrow linewidth monolithic WGLs with arbitrary laser wavelengths. The WGLs typically operated on two degenerate polarization modes but could be restricted to a single polarization mode through the use of polarised pump light. Given the flexibility of the femtosecond laser direct-write technique for creating WBGs on a per-period basis (where each grating period can be individually specified in position), there exists the potential to further improve device performance and create novel DFB designs by tailoring the WBG to annul pump induced thermal chirp or use sampled grating structures to create multiple laser output lines. Our investigations in this field are ongoing.\\

This work was produced with the assistance of the Australian Research Council under the ARC Centres of Excellence \& LIEF programs.



\end{document}